%% file: afb.tex
\documentclass[11pt]{article}
\usepackage{moriond,epsfig}

\def\Journal#1#2#3#4{{#1} {\bf #2}, #3 (#4)}

\def\NIMA{{\em Nucl. Instrum. Methods} A}

\def\PLB{{\em Phys. Lett.}  B}

\def\ZPC{{\em Z. Phys.} C}
\def\EPC{{\em Eur. Phys. J.} C}

\def\be{\begin{equation}}
\def\ee{\end{equation}}
\def\bea{\begin{eqnarray}}
\def\eea{\end{eqnarray}}

\include{macro}

\begin{document}
\vspace*{4cm}
\title{$\boldmath \Ab$ STATUS OF RESULTS}

\author{ V. Ciulli }

\address{Dipartimento di Fisica, Universit\`a di Firenze, INFN Sezione di Firenze, \\ via Sansone 1, I-50019 Sesto Fiorentino, Italy}
%\address{Department of Physics, Theoretical Physics, 1 Keble Road,\\
%Oxford OX1 3NP, England}

\maketitle\abstracts{ The status of results on forward-backward asymmetry
in $\bb$ decays is reviewed. A comparison of \LEP\ measurements, with
emphasis on the final \A\ measurement with leptons,  
and a critical discussion of average from heavy flavour electroweak
combination is presented.          
}

\section{Introduction}

The high efficiency and purity with which b-quarks can be tagged allow
precise measurements of the forward-backward asymmetry in $\bb$
decays, $\Ab$, defined as
\be
\Ab = \frac{\sigma_{\mathrm F}-\sigma_{\mathrm B}}{\sigma_{\mathrm F}+\sigma_{\mathrm B}} \ ,
\label{eq:ab}
\ee
where the cross sections are integrated over the full forward (F) and
backward (B) hemisphere. This asymmetry is the observable with the
highest sensitivity to $\sw$ at \LEP\ and thus an important test of the
Standard Model. 

Most experiments measure $\Ab$ from a fit to the differential cross
section with respect to the scattering angle: 
\be
\frac{1}{\sigma}\frac{d\sigma}{d\cos\theta}=\frac{3}{8}(1+\cos^2\theta)+\Ab\cos\theta \ .
\ee   
This is statistically more powerful than using Equation~\ref{eq:ab}
and is independent of non-uniform angular acceptance. The scattering
angle $\theta$ is measured using the thrust axis of the
event, oriented along the antiquark-quark direction.  
Therefore to measure the b forward-backward asymmetry the quark 
flavour needs to be tagged and the quark has to be separated from the
antiquark. Several methods have being exploited by the four \LEP\
collaborations. 

Leptons issued by direct $\bl$ decays provide both
flavour and charge tag. Thanks to the hard fragmentation and high mass
of b-quarks they are characterised by large total and transverse
momenta, $p$ and $\pt$, and their charge is correlated to the decaying
quark charge.  
However the quark charge at production is the relevant quantity for
the asymmetry measurement, so corrections are required for the effects
of $\bmix$ mixing.

Lifetime based taggings are also used to select high purity samples
of b-quarks, the quark charge then being estimated by jet- and
vertex-based charge taggings. 
   
Any analysis measures the asymmetry in the selected sample as given
by: 
\be
A_{FB}^{\mathrm meas} = \sum_\q (2\omega_q-1)\eta_q\mathrm{A}^\q_{\mathrm{FB}} ,
\ee
where $\eta_\q$ is the fraction of $\qq$ events in the sample and
$\omega_\q$ is the probability to correctly tag the quark charge.  
These are usually estimated on Monte Carlo simulated events and
depend on many parameters used to tune the simulation to the
real data. Some of these physics inputs are also measured using data
collected at the Z peak and therefore have common systematics 
between them and with the asymmetry. The four \LEP\ collaborations and
\SLD\ have agreed on a common list of these physical quantities and on
an averaging procedure to take into account correlations, resulting in
a simultaneous fit of the most important heavy flavour electroweak
observables~\cite{lephf}.

The only new result on $\Ab$ at this conference is the final
\A\ result with leptons~\cite{alep}, therefore emphasis is given here
to \LEP\ measurements using leptons. A review of the results obtained
with jet-charge based measurements can be found elsewhere~\cite{elsing}.    
The impact on the heavy flavour electroweak combination of the new \A\
measurement is also discussed in detail. 

\section{$\Ab$ measurement with leptons}

As pointed out before, leptons are a powerful tool for tagging b
quarks/antiquarks. However high momentum leptons are also produced in
$\cl$ decays, even if with a lower transverse momentum, as this is
limited by half the quark mass. Both b- and c-quarks decays in
electrons or muons with a branching ratios of about 
10\%. These leptons always have the same charge sign as the decaying
quark, but for charm decays an antifermion is produced out of a
fermion. Because of this fermion/antifermion flip, cascade decays
$\bcl$ have opposite forward-backward asymmetry with respect to
direct $\bl$ decays. The same holds for $\cl$ decays, since $\Ac$ has
the same sign as $\Ab$. Therefore the measured $\Ab$ shows a large sensitivity
to the sample composition. In addition the asymmetry in $\b$ events is also diluted
because of $\bmix$ mixing by a factor $1-2\chib$, where
\be
\chib= \chi_{\d} f_{\B^0_\d}  \frac{{\mathrm{BR}}(\B_\d^0\rightarrow
  \ell)}{{\mathrm{BR}}(\b\rightarrow \ell)} + \chi_{\s} f_{\B^0_\s}
\frac{{\mathrm{BR}}(\B_\s^0\rightarrow
  \ell)}{{\mathrm{BR}}(\b\rightarrow \ell)} 
\label{eq:mix}
\ee  
is the integrated mixing parameter ($ f_{\B^0_\d} $ and $ f_{\B^0_\s}
$ are the production fractions of $\B^0_\d$ and $\B^0_\s$). 

I will discuss in the following how the flavours and direct/cascade
decays are separated and how $\chib$ is measured from the data
themselves.

\subsection{Multivariate analysis} 

In an analysis based simply on lepton total and transverse
momentum, $\Ab$ can be measured on a sample enriched in $\bl$
decays by a cut on the lepton transverse momentum. As an example,
Table~\ref{tab:ptcut} shows the sample composition for 
the \A\ lepton sample, requiring a lepton transverse momentum greater
than $1.25$ GeV, and the correlation between the lepton charge and the
quark charge at decay time.
On the contrary $\cl$ decays can be only poorly separated on a 
statistical basis from $\bcl$ decays using these variables. 

\begin{table}[t]
\caption{Sample composition for $\pt>1.25$ GeV in the \A\ lepton
  sample. The correlation between the lepton charge and the decaying
  quark charge is also shown.\label{tab:ptcut}}
\vspace{0.4cm}
\begin{center}
\begin{tabular}{|l|c|c|}
\hline
Lepton source & fraction ($\pt>1.25$ GeV) & charge correlation \\
\hline
$\bl$,$\bccl$ & $0.795$ & 1  \\
$\bcl$ & $0.046$ & \hspace{-2ex}$-1$  \\
$\cl$ & $0.048$ & 1  \\
background & $0.111 $ & weak  \\
\hline
\end{tabular}
\end{center}
\end{table}

To enhance the sensitivity on both~\footnote{The determination of $\Ab$ also profit from a
better measurement of $\Ac$, because of its dependence
on it} $\Ab$ and $\Ac$, \A~\cite{alep}, \DELP~\cite{delphi}
and \OP~\cite{opal_jet,opal_up} use multivariate analyses, which allow
to improve the separation between the different contributions to the
lepton sample. 

\begin{figure}
\begin{center}
\begin{picture}(100,250)(0,0) 
\put(-190,0){\epsfig{figure=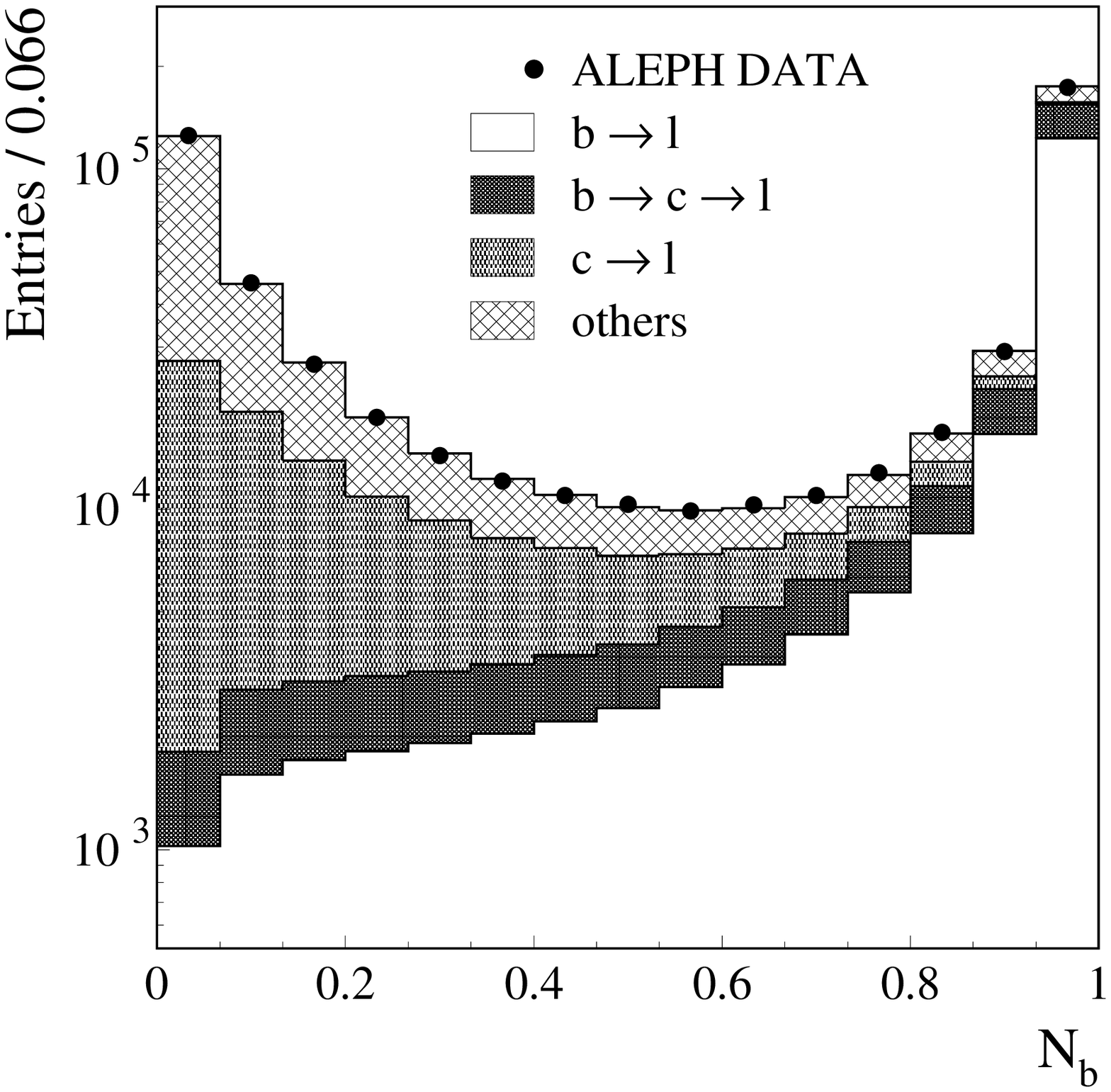,height=85mm,width=85mm}}
\put(50,0){\epsfig{figure=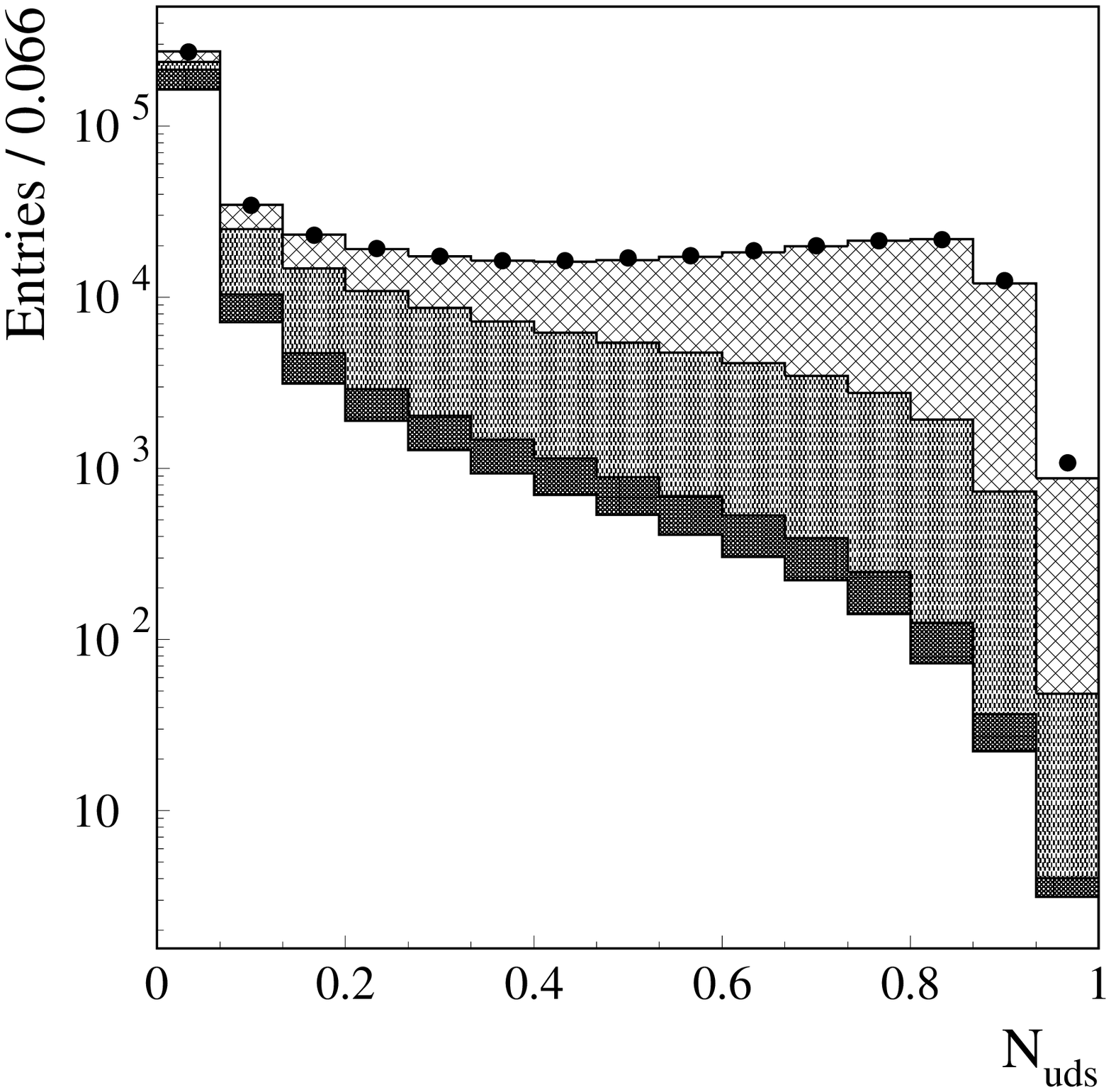,height=85mm,width=85mm}
}
\end{picture}
\end{center}
\vspace{-1em}
\caption{The distribution of the discriminating variables $N_{\b}$ 
and $N_{\ud}$ in the \A\ lepton sample.}
\label{fig:nb}
\end{figure}

To distinguish the event quark flavour, in addition to the lepton $p$
and $\pt$ \A\ uses information from: 
\begin{itemize}
\item lifetime tags based on the transverse impact parameter of
  tracks and on inclusively reconstructed secondary
  vertices;
\item missing energy due to neutrinos produced in semileptonic quark decays;
\item quark-mass related tags: sum of transverse momentum of
  tracks in the most energetic jet of the event and momentum of the highest
  momentum particle of the event;
\item a $\D^{*\pm}$ tag based on the soft pion issued in
  $\D^{*\pm}\to\D^0\pi^{\pm}$ decays. 
\end{itemize}
All these variables are combined in a neural network trained to
separate $\b$ from $\ud$ events~\footnote{Here $\ud$ events are
  referred all together as a single ``light-quark'' flavour, since in
  this context their properties do not differ significantly}. Because of
the limited acceptance of the vertex detector, the
discriminating power of the above variables (the lifetime tag ones in
particular) depends on the thrust $|\cos\theta|$ and therefore this
variable is also used in the neural network as a control variable. The
distributions of the two neural network outputs, called $N_\b$ and
$N_\ud$, are shown in Figure~\ref{fig:nb}. Since $\cc$ events
have intermediate properties a good discrimination of all the three
flavours is obtained in the plane ($N_\b$,$N_\ud$).

\begin{figure}
\begin{center}
\epsfig{figure=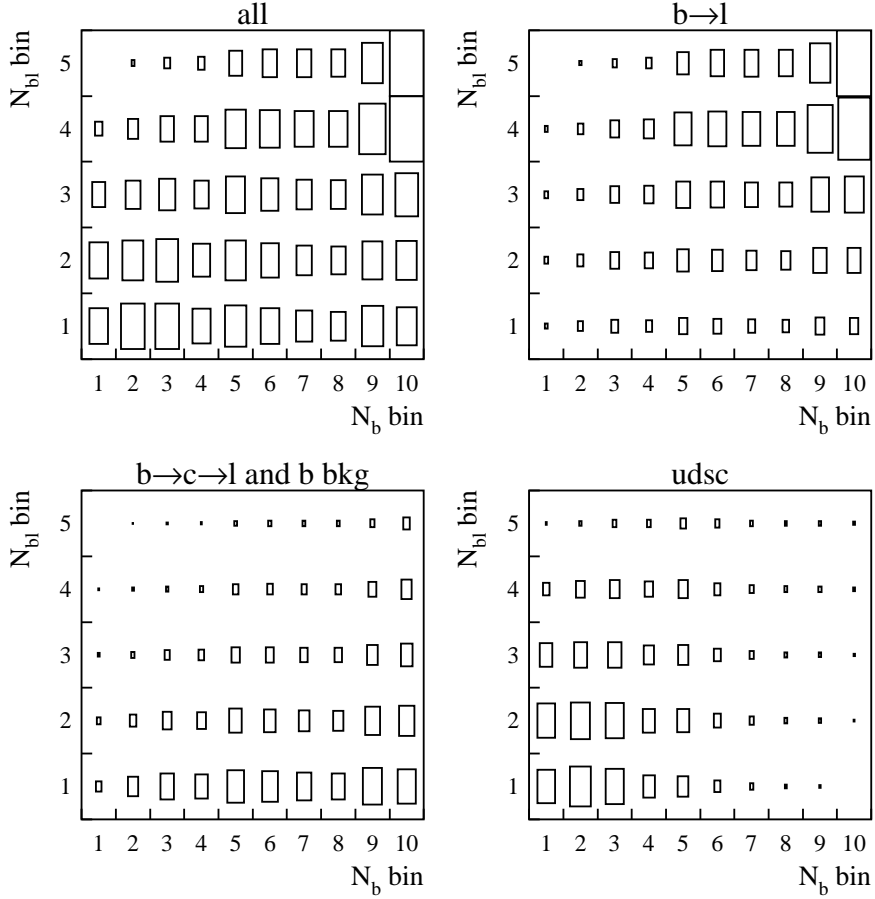,height=14cm}
\vspace{-2em}
\caption{Distribution in bins of $(N_{\b},N_{\b\l})$ in the \A\
  Monte Carlo simulation. 
\label{fig:binbl}}
\end{center}
\end{figure}

In order to separate $\bl$ and $\bcl$ decays the discriminating power
of lepton $p$ and $\pt$ is improved using the different properties of the
lepton jet in direct and cascade decays. Several variables are
built to distinguish whether in the jet, lepton excluded, only $\D$
decay products or both $\D$ and ``virtual'' 
$\W$ decay products are present. These variables are then combined
with the lepton kinematic properties in a neural network, which output
is called $N_{\mathrm{bl}}$. Figure~\ref{fig:binbl} shows the
distribution of the different processes in bins of $N_{\mathrm{bl}}$ versus
$N_\b$ in a simulated sample enriched in $\b$ and $\cq$ content by a cut
on the $N_{\ud}$ variable. The three main sources of leptons are
clearly well separated. The binning of the neural network outputs is
chosen as to ensure an almost equal occupancy of the different bins,
in order to minimise statistical fluctuations in the simulated
sample~\footnote{A Monte 
  Carlo sample of about 8 million hadronic Z decays is used, as well
  as two dedicated heavy flavour samples of 5 million $\bb$ decays and
  2.4 million $\cc$ decays. These figures must be compared with a full
  \LEP1\ statistics of about 4 millions hadronic Z decays. With the
  chosen binning the effect of statistical fluctuations in the
  simulated sample is found to be negligible.}.           

The use of lepton jet variables 
to enhance the separation of direct from cascade decays has been pioneered by
\OP~\cite{opal_jet,opal_up}. In their analysis two neural networks,
respectively for $\bl$ and $\cl$ decays, are built combining these
variables with lepton $p$ and $\pt$, a mass tag and two lifetime tags
(one based on the impact 
parameter of the tracks belonging to the lepton jet and the other most
energetic jet in the event, the other based on the lepton
impact parameter). The latest are actually used only as
anti-b tag in the $\cq$ neural network, since the $\b$ one is aimed at
separating $\bl$ from both lighter quarks decays and $\bcl$ decays. 
As a consequence in the plane defined by the two neural network
outputs, the $\bcl$ decays are not well separated from the background
due to $\ud$ and $\cq$ events, with some loss of information on $\Ab$.
On the other hand, the separation of $\cl$ decays from other sources
of leptons (either true or fake) should be optimal, since a dedicated
neural network is designed for this purpose.

\begin{figure}
\begin{center}
\begin{picture}(100,250)(0,0) 
\put(-190,0){\epsfig{figure=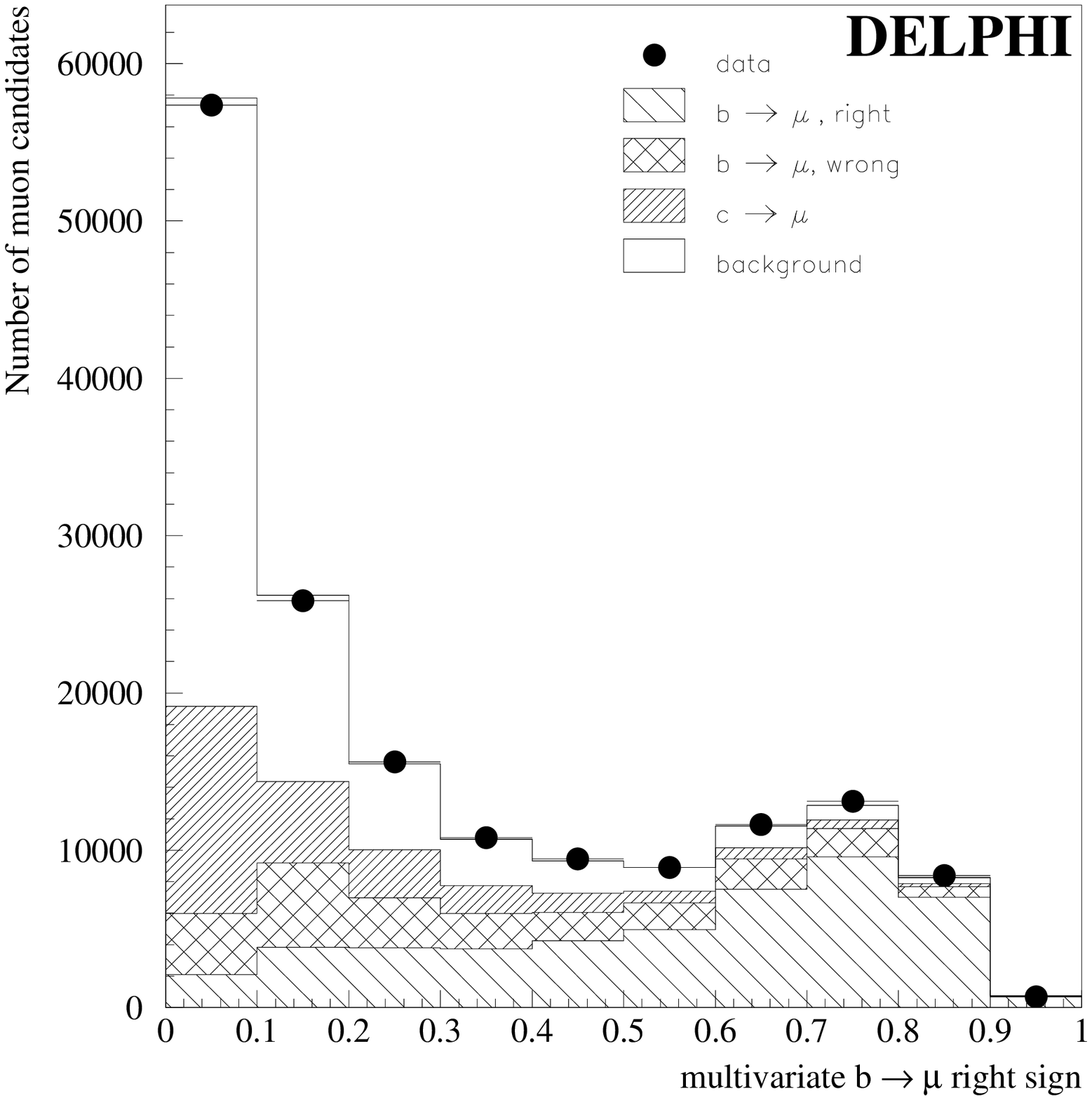,height=85mm,width=85mm}}
\put(50,0){\epsfig{figure=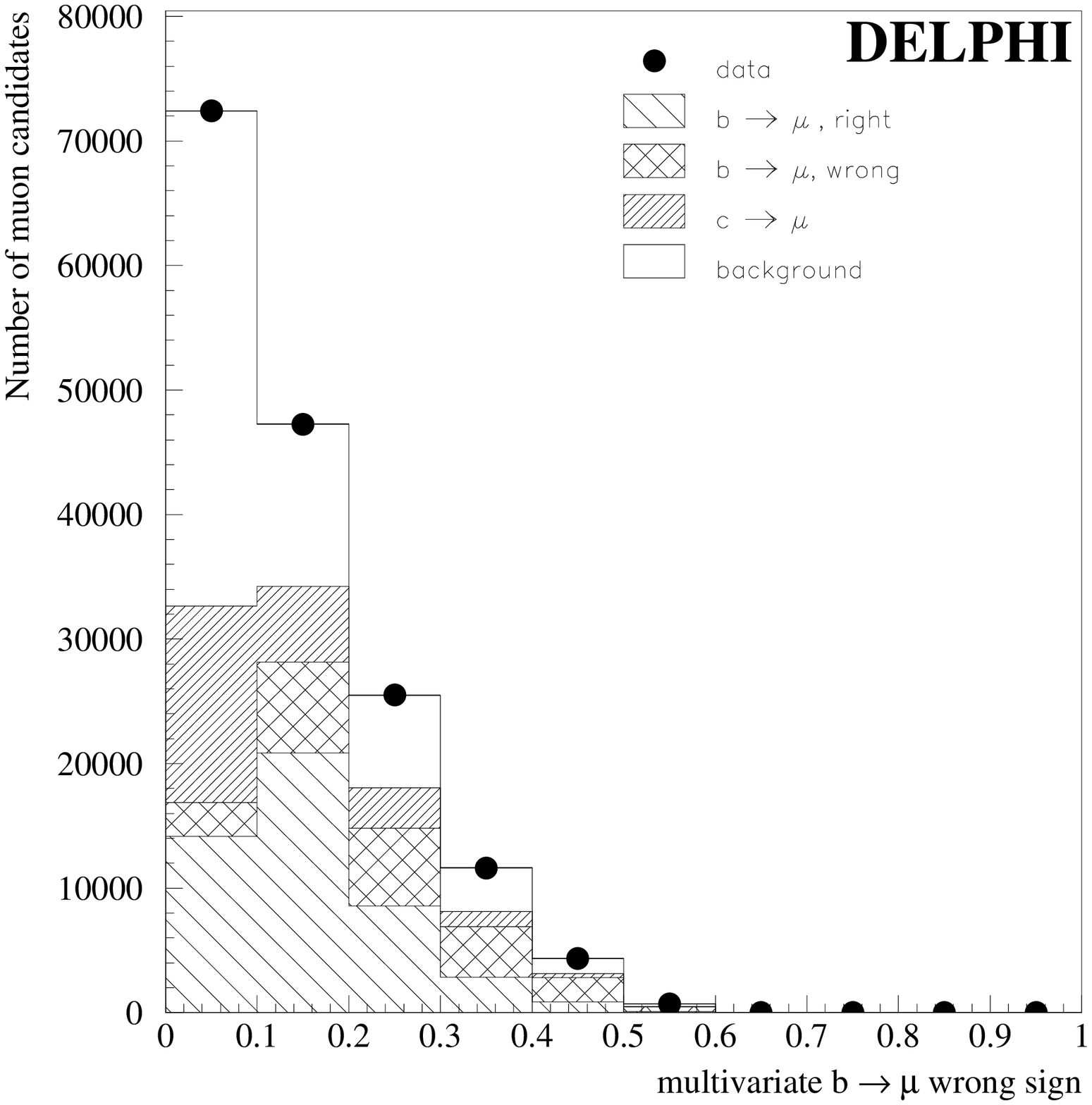,height=85mm,width=85mm}
}
\end{picture}
\end{center}
\vspace{-1em}
\caption{\small The distribution of the discriminating variables in
  the \DELP\ muon sample. Right (wrong) refers to the correct
  (incorrect) tagging of the b-quark charge at production time.}
\label{fig:delp}
\end{figure}

\DELP~\cite{delphi} used a lifetime tag to better separate b-quarks
and c-quarks and the correlation between the lepton charge and the
jet-charge in the opposite hemisphere to enhance the separation of
$\bl$ to $\bcl$ decays . This correlation is negative for $\bl$ decays and
positive for $\bcl$ decays. The sign is reverted  
if the lepton comes from a neutral B meson which undergoes 
$\bmix$ mixing. However in both cases using the jet-charge helps in tagging
the correct quark charge at production time. The draw-back is the
consequent large statistical and 
systematic correlation with the jet-charge based measurements.
Figure~\ref{fig:delp} shows the distributions of the discriminating
variables in the muon sample.         

\subsection{Mixing measurement}

The integrated mixing parameter $\chib$ (Equation~\ref{eq:mix}) is
measured using dileptons events. If both leptons
are produced by $\bl$ decays, the measured fraction of like-sign
dilepton events is proportional to $2\chib(1-\chib)$, allowing to
extract $\chib$. Background due to other processes and misidentified
hadrons has to be taken into account too, but requiring both leptons
to have a large $\pt$ or, even better, a ($\bl$)-like neural network
output, it is reduced at the 15\% level.

The new \A\ result for $\chib$ is 
$$
\chib = 0.1196\pm 0.0049\; \stat \;
{}^{\displaystyle{+0.0043}}_{\displaystyle{-0.0050}} \;\syst.
$$   
The largest contribution to the systematic error is due to the
modelling of semileptonic decays. It must be pointed out that the
previous \A\ result~\cite{alephpt}, $\chib=0.1246\pm 0.0051\; \stat \;
\pm 0.0052  
\;\syst $, would be $\chib=0.1193$ if the latest measured values
of the branching ratios  BR$(\bl)$, BR$(\bcl)$ and BR$(\cl)$ were taken
into account, therefore the new result is fully compatible with
the old one. 

\section{Results} 

\begin{figure}
\begin{center}
%\begin{picture}(100,400)(0,0) 
%\put(-200,0){
\epsfig{figure=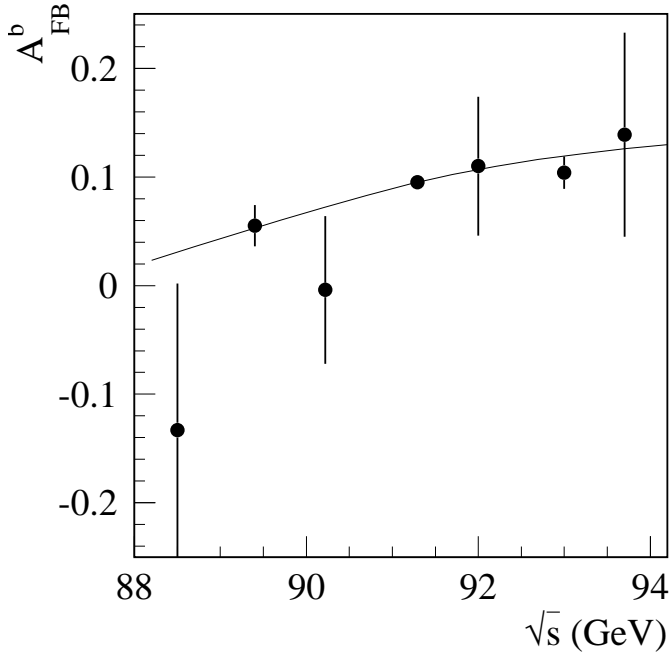,height=10cm}
%}
%\end{picture}
\vspace{-2em}
\caption{$\Ab$ measurements as a function of the
  centre-of-mass  energy. The curve is the Standard Model expectation
  fitted to the measured values of the asymmetry.
\label{fig:afb}}
\end{center}
\end{figure}

The asymmetry is measured separately at each centre-of-mass energy
point. The final \A\ results are shown in Figure~\ref{fig:afb}. 
These asymmetry measurements performed at peak and off-peak energies are
extrapolated to $\mathrm{M_Z}$. Then the QED, QCD and $\Z-\gamma$
interference corrections are applied~\cite{corr} to obtain
the pole asymmetry:   
$$
\Abp = 0.0998 \pm 0.0040 \; \stat \; \pm 0.0017 \; \syst \ . 
$$
Systematic uncertainties are shown in Table~\ref{tab:aleph}. The main
systematic error is due to the uncertainty on the mixing
parameter. This is mostly  
statistical, since the sources of systematic error are common and
partially cancel out.

\begin{table}[t]
\caption{Systematic uncertainties on the \A\ final result. Numbers are
  given in units of $10^{-2}$ \label{tab:aleph}}
\vspace{0.4cm}
\begin{center}
\begin{tabular}{|l|c|}
\hline
Error sources & $\Delta(\Ab)$ \\
\hline
Semileptonic branching ratios & $0.034$ \\
Lepton modelling & $0.090$ \\
Detector simulation & $0.015$ \\
Background asymmetries & $0.002$ \\
B and D physics & $0.032$ \\
Mixing & $0.132$ \\
\hline
Total & $0.169$ \\
\hline
\end{tabular}
\end{center}
\end{table} 

\begin{figure}[t]
\begin{center}
\epsfig{figure=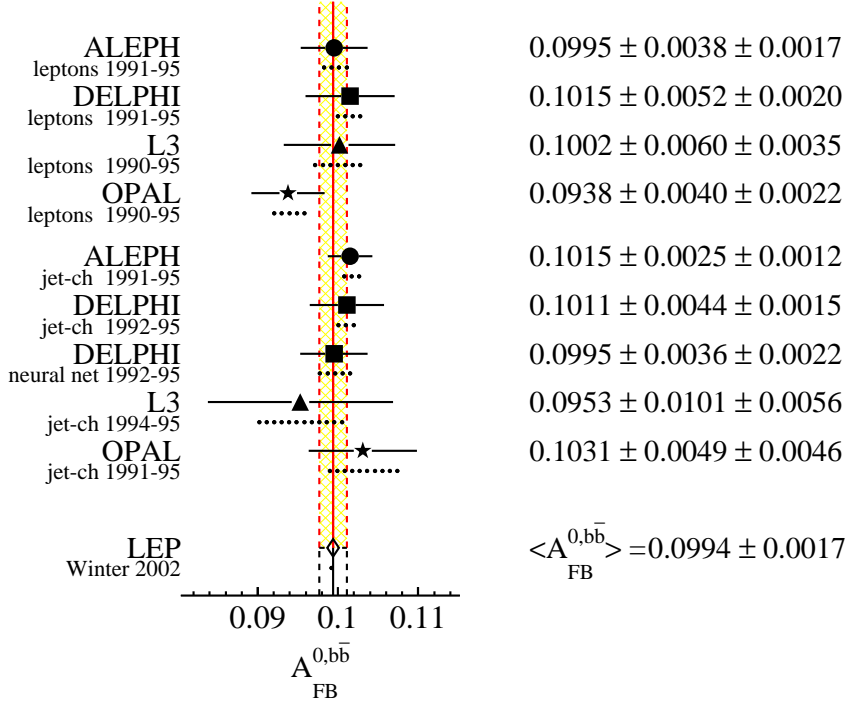,height=16cm}
\vspace{-13em}
\caption{$\Abp$ measurements used in the heavy flavour
  combination. The $\Abp$ measurements with D-mesons contribute only
  very little weight and are not shown in the plot. The values shown are
corrected to the same input parameters (including 
the other fitted observables which are fixed to the results of the full fit).  \label{fig:all}}
\end{center}
\end{figure}

This measurement is combined with the inclusive \A\ result and with the
results of the other \LEP\ experiments~\cite{lephf}. The
$\b$ forward-backward asymmetry is fitted together with the most  
relevant electroweak heavy-flavour observables and with the most
important input parameters, taking into account the common systematics
and the statistical correlation. Figure~\ref{fig:all} shows the most
precise $\Abp$ measurements included in the fit. 
The combination yields 
$$
\Abp = 0.0994 \pm 0.0017 \ .
$$
The main error source is statistics, and the small systematic
error is mostly uncorrelated between the different experiments, as
shown in Table~\ref{tab:all}. All results, either using leptons or
inclusive methods, are consistent between the four \LEP\ collaborations.

\begin{table}[t]
\caption{Sources of uncertainty on $\Abp$ average. Numbers are
  given in units of $10^{-2}$ \label{tab:all}}
\vspace{0.4cm}
\begin{center}
\begin{tabular}{|l|c|}
\hline
Error sources & $\Delta(\Abp)$ \\
\hline
Statistics & $0.16$ \\
\hline
Internal systematics & $0.06$ \\
Common systematics & $0.04$ \\
Total Systematics & $0.07$ \\
\hline
Total & $0.17$ \\
\hline
\end{tabular}
\end{center}
\end{table} 

With respect to the summer 2001 preliminary result~\cite{summer2001},
$\Abp=0.0990 \pm 0.0017$, the average has changed by a quarter of a sigma
because of the final \A\ result. As we have shown before, this change
is not due to the new measurement of the mixing parameter
$\chib$~\footnote{The combination procedure already took into account the
  changes in the branching ratios values with respect to the time the
  previous $\chib$ measurement was published, so the $\chib$ value 
  used in the fit is unchanged with respect to summer 2001.}. The \A\
measured $\Ab$ and to a less extent the other \LEP\ measurements, have
changed because of the final \A\ measurement of $\Ac$. With respect to
the preliminary result~\cite{summer2001}, off-peak energies measurements
have been performed and a problem due the too large number of
bins in which the analysis was performed has been fixed. 
This measurement, $\Acp=0.0732 \pm 0.0053 \; \stat \; \pm 0.0037 \;
\syst$, is today the most precise \LEP\ result on $\Acp$.      

\section{Conclusions}

The status of $\Ab$ measurements has been discussed with particular
emphasis on the final \A\ measurement with leptons 
$$
\Abp = 0.0998 \pm 0.0040 \; \stat \; \pm 0.0017 \; \syst \ . 
$$
Using this new result the heavy flavour electroweak combination yields
$$
\Abp = 0.0994 \pm 0.0017 \ .
$$
All results, either using leptons or inclusive methods, are consistent between
the four \LEP\ collaborations. 

\section*{Acknowledgements}
I am grateful to the \LEP\ Electroweak Heavy Flavour Working Group
and in particular to R.~Tenchini and E.~M.~P.~Antilogus
for their help in preparing this talk. I also own a special thank to
D.~Abbaneo.

\end{document}

%% file: macro.tex
\def \etal  {\hbox{\it et al.}}
\def \l    {{\mathrm{l}}}

\def \d    {{\mathrm{d}}}
\def \s    {{\mathrm{s}}}
\def \cq    {{\mathrm{c}}}
\def \b    {{\mathrm{b}}}

\def \q    {{\mathrm{q}}}
\def \D    {{\mathrm{D}}}
\def \B    {{\mathrm{B}}}

\def \Z    {{\mathrm{Z}}}
\def \W    {{\mathrm{W}}}

\def \ud   {{\mathrm{uds}}}

\def \stat {{\mathrm{(stat.)}}}
\def \syst {{\mathrm{(syst.)}}}

\def \pt   {p_{\perp}}
\def \LEP    {{\sc Lep}}
\def \CERN   {{\sc Cern}}

\def \A      {{\sc Aleph}}
\def \SLD    {{\sc Sld}}
\def \OP     {{\sc Opal}}
\def \DELP   {{\sc Delphi}}

\newcommand {\bl} {\b \to \ell}
\newcommand {\cl} {\cq \to \ell}

\newcommand {\bcl} {\b \to \cq \to \ell}

\newcommand {\bccl} {\b \to \bar \cq \to \ell}

\def \bb    {\Z \to \b \bar{\b}}
\def \cc    {\Z \to \cq \bar{\cq}}
\def \qq    {\Z \to \q \bar{\q}}

\def \Ab    {{\mathrm{A}}^\b_{\mathrm{FB}}} 
\def \Ac    {{\mathrm{A}}^\cq_{\mathrm{FB}}} 
\def \Abp   {{\mathrm{A}}^{0,\b}_{\mathrm{FB}}} 
\def \Acp   {{\mathrm{A}}^{0,\cq}_{\mathrm{FB}}}

\def \sw    {{\mathrm{sin}}^2\theta^{\mathrm{eff}}_{\mathrm{W}}}

\def \chib  {\bar \chi} 
 
\def \bmix  {\B^0{\mathrm{-}}\bar \B^0}

\def \rightdownarrow
 {\kern.5em
 \rule[.5ex]{.15mm}{2ex}
 {\mbox{$\kern-0.1em{\rightarrow}$}}}